\documentclass[preprint,aps,draft]{revtex4}

\usepackage{graphicx}
\usepackage{dcolumn}
\usepackage{bm}

\begin{document}


\title{Tunneling: From Milliseconds to Attoseconds}

\author{G\"unter Nimtz}
 \altaffiliation
 {G.Nimtz@uni-koeln.de}
\affiliation{
II. Physikalisches Institut, Universt\"at zu K\"oln \\
Z\"ulpicherstrasse 77, 50937 K\"oln
}%


\date{\today}

\begin{abstract}

How much time does a wave packet spent in tunneling a barrier?
Quantum mechanical calculations result in zero time inside a barrier.
In the nineties analogous tunneling experiments with microwaves were carried out confirming quantum mechanics.
Electron tunneling time is hard to measure being extremely short. However, quite recently the atomic ionization tunneling time has been measured.
Experimental data of photonic, phononic, and electronic tunneling time is available now. It appears that the tunneling time is a universal property in the investigated time range of twelve orders of magnitude.
\end{abstract}

\maketitle

Several theoretical investigations resulted in zero time tunneling, i.e. the time spent inside a barrier is expected to be zero~\cite{Hartman,Carniglia,Ali,Low,Wang}. Experimental studies with microwaves have shown that the measured short barrier traversal time $\tau$ is spent at the front boundary of a barrier, whereas actually zero time is spent inside the barrier~\cite{Nimtz}. The measured various photonic data are pointing to a universal time behavior, which is given by
\begin{eqnarray}
\tau & \approx & \frac{1}{\nu} = T \\
\tau & \approx & \frac{h}{E} ,
\end{eqnarray}
where $\tau$ is the measured barrier traversal time, $\nu$ the carrier frequency and T is the oscillation time of the electromagnetic wave packet.
h is the Planck constant and E is the particle energy in the case of a wave packet with a rest mass.

This empirical universal tunneling time behavior was studied theoretically by Esposito~\cite{Esposito}. He obtained the modified relation
\begin{eqnarray}
\tau_A = \frac{1}{\nu} \cdot A, \label{A} \label{Espo}
\end{eqnarray}
where A is depending on the special barrier and wave packet in question.

Later experimental data on phonon (acoustic) tunneling have pointed to a similar barrier traversal time~\cite{NimtzA}.
Recently, extremely short electronic tunneling data in the atto seconds range were observed in the ionization process of Helium \cite{Keller}, which also fit in this approximate universal tunneling time relation of Eq.\ref{Espo}.

\begin{table*}
\begin{center}
\begin{tabular}{|l|l|c|c|c|}

\hline
\hline Tunneling barriers & Reference  & $\tau$   &  $T=1/\nu$ & $\tau_A$  \\
\hline
\hline {\it Frustrated total reflection}   & Haibel/Nimtz & 117\,ps & 120\,ps & 81\,ps \\
\hline
\cline{2-4}   & Balcou/Dutriaux  &  30\,fs &  11.3\,fs & 36.8\,fs  \\
\hline
\cline{2-4}   & Mugnai et al. & 134\,ps & 100\,ps  & 87\,ps  \\
\cline{2-4}
\hline
\hline {\it Photonic lattice} & Steinberg et al. & 2.13\,fs & 2.34\,fs & 2.02\,fs \\
\hline
\cline{2-4} &  Spielmann et al.  &  2.7\,fs  &  2.7\,fs & 2.98\,fs  \\
\hline
\hline {\it Undersized waveguide} & Enders/Nimtz & 130\,ps & 115\,ps & 128\,ps \\
\hline
\hline {\it Ionization tunneling} & Eckle et al.  & 6\,as & 75\,as & 4.25\,as\\
\hline
\hline {\it Acoustic (phonon) tunneling } & Yang et al. & 0.8\,$\mu$s &  1\,$\mu$s &  0.6$\mu$s \\
\hline
\hline
\end{tabular}
\end{center}
\caption{Tunneling times $\tau$, $1/\nu$ = T, and $\tau_A$. References of the data are given below and in \cite{NimtzA}}
\end{table*}

The barrier traversal time of photons and phonons was obtained either by directly measuring the barrier traversal time or by calculating the time by the phase time approach
\begin{eqnarray}
\tau = - d\phi/d\omega,\label{phase}
\end{eqnarray}\\
where $\phi$ is the phase shift of the wave and $\omega$ is the
angular frequency; $\phi$ is given by the real part of the wave
number k times the distance x. In the case of evanescent modes
and of tunneling solutions  the wave number k is purely imaginary. Thus propagation of
tunneling modes appears to take place in zero time \cite{Low}.

An extremely short atom ionization electron tunneling time was measured in a sophisticated experiment ~\cite{Keller}. Actually, this time does fit the modified universal tunneling time concept as shown in the Table. (It is assumed E = 54.39 eV and V$_0$ = 78.98 eV neglecting the applied laser field of about 10$^{-14}$ W/cm$^2$.)
The factor A is given for the case of a tunneling particle described by the Schr\"odinger equation as follows
\begin{eqnarray}
\tau_A = \frac{\hbar}{ \sqrt{E(V_0 - E)} }= \frac{1}{\nu} \cdot \frac{E}{4 \pi^2 (V_0 - E)},
\end{eqnarray}
where E and $V_0$ are the particle energy and the height of the square potential barrier.

Hartman's early quantum mechanical study on tunneling of a wave packet revealed
 a barrier traversal time independent of barrier length and zero time inside barriers~\cite{Hartman}. The calculations
were confirmed by above mentioned photonic tunneling experiments, see for instance~Ref.\cite{Nimtz}.
The barrier traversal time
seems to be in first order approximation universal for photons, phonons, and electrons in the investigated time range of 12 orders of magnitude.


\end{document}